\documentstyle[prl,aps,floats,twocolumn]{revtex}
\begin{document}
\title{A two-qubit algorithm involving quantum entanglement}
\author{Arvind$^1$\thanks{arvind@physics.iisc.ernet.in} and
N. Mukunda$^2$\thanks{nmukunda@cts.iisc.ernet.in}\thanks{Also at 
Jawaharlal Nehru Centre for Advanced Scientific Research,
Jakkur, Bangalore 560 064, India}}
\address{$^1$Department of Physics, 
Guru Nanak Dev University, Amritsar 143 005, India }
\address{$^2$Center for Theoretical Studies and 
Department of Physics, Indian
Institute of Science, Bangalore 560012, India }
\date{\today}
\maketitle
\draft
\begin{abstract}
The phenomenon of quantum entanglement is fundamental to the 
implementation of quantum computation, and requires at least two 
qubits for its demonstration. However, both Deutsch algorithm and 
Grover's search algorithm for two bits  do not use entanglement.  
We develop a Deutsch-like problem, where we consider all possible 
binary functions for two bit inputs and distinguish their even or 
odd nature. The quantum algorithm to solve this problem requires 
entanglement at the level of two qubits. The final solution suggests 
that an NMR implementation of the problem would lead to interesting 
results.
\end{abstract}
The fusion of ideas from classical information theory and the 
foundations of quantum mechanics has led to a quiet revolution 
in computer science today - namely, quantum computation~\cite{benn-nat-00}. 
One can now envisage building computing devices to physically
implement fast quantum algorithms. Quantum algorithms rely on 
truly quantum phenomena like entanglement, interference and 
quantum superposition to achieve a significant computational 
advantage over classical 
algorithms~\cite{divin-sc-95,lloyd-sc-95,chua-sc-95,pres-proc-98,knill-nat-00}. 
The algorithms designed thus far to demonstrate the power of quantum computing 
range from the Deutsch-Jozsa algorithm to evaluate the 
constant or balanced nature of a function~\cite{deu-roy-92} and Grover's
quantum search algorithm~\cite{grover-prl-97} which achieve a polynomial
speed-up, to Shor's factorization algorithm which
gains an exponential advantage over its best known
classical counterpart~\cite{shor-siam-97}.

Quantum entanglement shows up qualitatively at the level of 
two qubits. It can be visualised as the non-separability 
of the state of a composite quantum system into the states of 
its parts~\cite{ekert-amj-94}. 
For example, if two qubits are in a state such as
$\frac{1}{\sqrt{2}}\{\vert 01 \rangle - \vert 10 \rangle \}\/$,
which is not resolvable into the tensor product of
the states of the individual qubits, they are entangled.
Neither qubit by itself has a definite pure state, in contrast to a 
classical system which can be completely resolved into the states of
each part of the system. The problem of quantum entanglement
for general (pure or mixed) states has attracted a lot of
attention~\cite{peres-prl-96,horo-prl-97}.

It is quantum entanglement which prevents the mapping or realisation
of a quantum computation using just  classical waves.  Consider the 
polarisation states of a classical light beam. These states are in 
one-to-one correspondence with the states of a qubit. All possible states 
can be realised by using one half-wave and two quarter-wave 
plates. One can pass from any one chosen polarisation state to all 
others in this way, i.e. all $U(2)\/$
transformations can be implemented using these gadgets~\cite{simon-pla-90}.
Therefore a single qubit has a classical analogue.  On the other hand, 
it is not possible to map the states of a two-qubit system onto the polarisation
states of two light beams. The entangled states of the two qubits have no
classical counterpart. Therefore at the level of two qubits itself
the possibility of mapping a quantum computer onto 
classical optical fields breaks down\cite{arv-cal-00}.

The Deutsch-Jozsa(DJ) algorithm was the first and the simplest example
of an algorithm that demonstrated the appreciable advantage of
quantum computing~\cite{deu-roy-92,cleve-royal-98}. However, 
it was realised recently 
that this algorithm requires entangling transformations only for three  or
more input bits.  For the case of two input qubits, 
it can be mapped onto an essentially classical optical 
problem~\cite{collins-pra-98,arv-qph-67,arv-cal-00}.
Generalisations of the DJ problem are also being
considered~\cite{dong-qph-59}.
Our aim in this paper is to discuss a simple two qubit problem which requires 
manipulation of  entangled states  for its solution in an essential way. 
We thus evaluate the global property of a function (its even
or odd nature) using fewer function calls than a classical algorithm. The algorithm
is designed to exploit the entangled states of the two qubits.
In the process of the formulation and solution of this problem 
we arrive at some interesting consequences for distinguishing non-orthogonal
states through measurement. It turns out that this is an issue relevant 
to quantum computation using NMR which has been the most successful 
technique to implement quantum computation schemes, including the DJ 
problem~\cite{cory-proc,gersh-sc-97,ch-nature-98,jones-jcp-98,lin-cpl-98,kavita-pra-00}.

Consider a Boolean function  defined from a two-bit domain space to a one-bit
range space: $f(x) : \{ 0,1 \}^{2} \rightarrow \{ 0,1 \}\/$.
There are four possible input values $(00),(01),(10)$
and $(11)$ and the output for each of these could be either $0\/$ or $1\/$.
There are thus 16 functions in all.   For a given function, the output
can have either: all ones, three ones and a zero, two ones and two zeros, 
three zeros and one one  or all zeros.  We can divide the function into  classes
$[0,4], \, [1,3], \,[2,2], \,[3,1], $ and \,$ [4,0]\/$, the first entry 
indicating the number of ones and the second indicating the number of 
zeros in the output.  The functions with an even number $(0,2,4)\/$ of ones 
(i.e. the functions $[0,4], \, [2,2]\/$ and $[4,0]\/$)  are defined as 
``{\bf Even}'' functions while the functions with an odd number $(1,3)\/$ 
of ones in the output (i.e. the $[1,3]$ and $[3,1]$ functions) are defined to 
be  ``{\bf Odd}'' functions. Using this evaluation criterion, of the 16 possible 
functions for the two-bit case, eight are even and eight are odd.

The appropriate question to be asked is: 
{\it \bf given a function, how to decide whether it is even or odd.}
Classically, the classification of a given function would require 
computing the function at all input points, since even the last 
output can change the class of a function.
We give here an algorithm which classifies a given function using
fewer function calls. What we require in our analysis is a gate
to call the function, and a judicious use of Hadamard transformations.
A Hadamard transformation mixes the two eigenstates of a qubit maximally.
\begin{eqnarray}
&
\begin{array}{c}
\vert 0\rangle \stackrel{H}{\rightarrow} 
\frac{\scriptstyle 1}{\scriptstyle \sqrt{2}}
(\vert 0 \rangle + \vert 1 \rangle) \\
\vert 1\rangle \stackrel{H}{\rightarrow} 
\frac{\scriptstyle 1}{\scriptstyle \sqrt{2}}
(\vert 0 \rangle - \vert 1 \rangle)
\end{array}
;\,
H = H^{-1}=\frac{1}{\sqrt{2}}
\left(\begin{array}{lr} 
{ 1} & {1}\\
{ 1} & {-1}
\end{array}
\right)  &
\end{eqnarray}
There are three different types of Hadamard transformations possible for the
two qubit system. One can either apply the Hadamard transformation 
selectively on the first or the second qubit, or non-selectively on
both the qubits. Explicitly,
\begin{eqnarray}
 H^1 &=& H \otimes I \nonumber \\
 H^2 &=& I \otimes H \nonumber \\
H^{1\mbox{-}2} &=&H \otimes H 
\label{hadamard}
\end{eqnarray}
where $1\/$ and $2\/$ label the qubit involved.

The function call mechanism is similar to the one used by the 
Deutsch problem~\cite{collins-pra-98,arv-qph-67}. 
Each function $f\/$ can be encoded by a unitary
transformation $U_f\/$, 
with its action on the eigenstates of the two qubits
being defined as
\begin{eqnarray}
&\vert x \rangle_{\mbox{\tiny 2-bit}}
\stackrel{U_{f}}{\longrightarrow}
(-1)^{f(x)} \vert x \rangle_{\mbox{\tiny 2-bit}}&
\nonumber \\
\nonumber \\
&
U_f=\left(
\begin{array}{cccc}
(-1)^{f(00)}&0&0&0\\
0&(-1)^{f(01)}&0&0\\
0&0&(-1)^{f(10)}&0\\
0&0&0&(-1)^{f(11)}
\end{array}
 \right)
&
\end{eqnarray}

There are sixteen $U_f\/$ matrices in all, with half of them 
being not separable. For example, consider the matrix with diagonal entries
$[1,1,1,-1]\/$; it cannot be written as a tensor product of two 
matrices, one belonging to each qubit. Therefore the transformations
$U_f\/$ are in general entangling in character. 
The sub-class of functions that are either constant or balanced in
the sense of Deutsch problem i.e. the functions $(4,0), (0,4)\/$ and $(2,2)\/$
are all separable in character. For example consider the matrix for a 
balanced function with diagonal entries $[1,1,-1,-1]\/$:  is actually a 
tensor product of two $2\times 2\/$  matrices with diagonal entries 
$[1,-1]\/$ and $[1,1]\/$. Therefore, the two-bit Deutsch
problem, namely distinguishing between constant and balanced functions, 
can be implemented using non-entangling transformations alone.
One can even conceive of using classical waves for its implementation as
only the concepts of superposition and interference are required for its solution.
The present problem of distinguishing between even and odd functions 
on the other hand requires entangling
transformations for its implementation. These entangling transformations can
produce entangled states which  do not have any analogue in the
classical world even given superposition and interference.

\begin{table}
\hspace*{-12pt}
\renewcommand{\arraystretch}{1.4}
\begin{tabular}{|p{0pt}cp{0pt}|p{0pt}cp{0pt}|p{0pt}cp{0pt}|p{0pt}cp{0pt}|p{0pt}cp{0pt}|}
\hline
&Class &&& Number &&& Nature &&& $U_f$ &&& DJ Class&\\
\hline
\mbox{}
&[0,4] &&& 1 &&& Even &&& Separable  &&& Constant&\\
\hline
\mbox{}
&[1,3] &&& 4 &&& Odd  &&& Entangling &&& ------  &\\
\hline
\mbox{}
&[2,2] &&& 6 &&& Even &&& Separable  &&& Balanced &\\
\hline
\mbox{}
&[3,1] &&& 4 &&& Odd  &&& Entangling &&& ------ &  \\
\hline
\mbox{}
&[4,0] &&& 1 &&& Even &&& Separable  &&& Constant&\\
\hline
\end{tabular}
\vspace*{12pt}
\caption{Characteristics of different classes of functions. In each class we give
number of functions, their even or odd nature, the entangling or 
separable nature of $U_f\/$
and their status in DJ problem.} 
\end{table}

The computation proceeds with both qubits 
initially in the state $\vert 0 0\rangle\/$.
The sequence of steps followed then is:
Apply the transformation $H^{1\mbox{-}2}\/$,
call the function by applying $U_f\/$, 
apply the selective Hadamard transformation $H^2\/$ 
on second qubit alone, call the function 
a second time through $U_f\/$, and 
finally again apply the two-bit Hadamard transformation
$H^{1\mbox{-}2}\/$. This leads to the result
\begin{eqnarray}
&&H^{1\mbox{-}2}\, U_f\, H^{2}\, U_f\, H^{1\mbox{-}2}\, \vert 00 \rangle 
\nonumber \\
&&\quad\quad=\frac{1}{2\sqrt{2}}\left[
\left( 
(-1)^{f(00)\oplus f(01)}
+(-1)^{f(10)\oplus f(11)}
\right)
\vert 0 0 \rangle \right.
\nonumber\\
&&\quad\quad\quad\quad+\,2\,
\vert 0 1 \rangle 
\nonumber\\
&&\quad\quad\quad\quad+ 
\left.
\left(
(-1)^{f(00)\oplus f(01)}
-(-1)^{f(10)\oplus f(11)}
\right)
\vert 1 0 \rangle 
\right]
\end{eqnarray}
For an ``even'' function the final state becomes 
$$
\frac{1}{\sqrt{2}}
\left(
\pm\vert 0 0 \rangle 
+\vert 0 1 \rangle 
\right)
$$
and for an ``odd'' function it becomes
$$
\frac{1}{\sqrt{2}}
\left(
\pm\vert 1 0 \rangle 
+\vert 0 1 \rangle 
\right)
$$
These final states are clearly different and hence can be
used to classify the function as ``even'' or ``odd''. 
However, unlike the case of the DJ algorithm, these states are not 
orthogonal.  Can one hence  unambiguously conclude from
a single  measurement the character of the function?
In conventional quantum measurement theory  one expects that, to distinguish
between such states the experiment has to be repeated at least a few times.
Recently there have been refinements where a single
measurement can unambiguously distinguish between such states, though
the experiment may not work all the time~\cite{huttner-pra-1996}.

The most successful method of implementing quantum algorithms
to date has been NMR. As far as NMR is concerned the 
measurements obtained from the above two states leads to
very different spectra~\cite{ernst-book}.  
To see it more clearly we compute the density 
matrices corresponding to the  
two non-orthogonal final states which are 
\begin{equation}
\rho_{\rm even} = \frac{1}{2}\left(
\begin{array}{cccc}
1&\pm 1&0&0\\
\pm1&1&0&0\\
0&0&0&0\\
0&0&0&0
\end{array}
\right), \, 
\rho_{\rm odd} = \frac{1}{2}\left(
\begin{array}{cccc}
0&0&0&0\\
0&1&\pm1&0\\
0&\pm1&1&0\\
0&0&0&0
\end{array}
\right)
\end{equation}
The  density matrix $\rho_{\rm even}\/$ 
has off-diagonal terms corresponding to
single quantum coherences and therefore
will give rise to a  line in the NMR spectrum. On the other hand
$\rho_{\rm odd}\/$ has off-diagonal terms corresponding to  
zero quantum coherence which does not give rise to any
observable NMR signal~\cite{ernst-book}. 

To further elaborate this point we calculate the reduced density
matrix corresponding to the second spin for both the cases.
\begin{eqnarray}
\rho^{(2)}_{\rm even} &=& \frac{1}{2}
\left(
\begin{array}{cc}
1&\pm1\\
\pm1&1
\end{array} 
\right)
= 
\frac{1}{2}
I +
\frac{1}{2}
\left(
\begin{array}{cc}
0&\pm1\\
\pm1&0
\end{array}
\right)  
\nonumber \\
\rho^{(2)}_{\rm odd} &=& \frac{1}{2}
\left(
\begin{array}{cc}
1&0\\
0&1
\end{array}
\right) =
\frac{1}{2}
I + 
\left(
\begin{array}{cc}
0&0\\
0&0
\end{array} 
\right)
\label{reduced}
\end{eqnarray}
Multiples of identity do not give rise to any NMR signal at  all.
After a multiple of identity is taken out, 
the two reduced density matrices are clearly very different.
The first one is not zero and has single quantum coherences,
therefore it 
will give rise to a line in the NMR spectrum while the second one is
zero and hence will lead to no measurable NMR signal.
Therefore a clear demarcation of 
even and odd functions is possible using the NMR quantum computer.
It might be
interesting to do experiments in this direction!
We can argue similarly for the first spin by calculating  reduced 
density matrices corresponding to it. It turns out that measurements
on the first spin alone do not distinguish between the
even or odd nature of the function.

The reduced density matrices~(\ref{reduced}) also show that for odd
functions we have quantum entanglement. It follows from the fact that
$\left[\rho^{(2)}_{\rm odd}\right]^2\,\not=\,\rho^{(2)}_{\rm odd}$. Therefore,
the second qubit alone, after partial trace is in a mixed state (in fact the 
reduced density matrix for the second qubit for odd functions is a 
multiple of identity and is therefore maximally mixed) indicating 
clearly, that the original two-qubit pure density matrix has entanglement
for odd functions.

We have described an algorithm for two qubits which  requires the 
implementation of entangling transformations for its execution. 
As explained by Schroedinger, entanglement is not just one way but {\em the}
way in which quantum mechanics differs from classical physics~\cite{sch-proc-35}.
The problem
we have addressed is a natural generalisation of the Deutsch problem. In our
solution we require two function calls as opposed to four for the classical
solution. Since in this problem the manipulation of entanglement is essential
it will be very interesting to implement this simple algorithm experimentally
and track down the amount of entanglement. This is being currently pursued
and will be reported elsewhere.
\vspace*{12pt}

\end{document}